# Development and Validation of a Tokamak Skin Effect Transformer model

J.A. Romero, J.-M.Moret[1], S.Coda[1], F.Felici[1], I. Garrido[2]

Laboratorio Nacional de Fusión, Asociación EURATOM/CIEMAT, 28040, Madrid, Spain.
[1] Centre de Recherches en Physique des Plasmas, Association EURATOM—Confédération Suisse, Ecole Polytechnique Fédérale de Lausanne, CRPP—EPFL, CH-1015 Lausanne, Switzerland
[2] Euskal Herriko Unibertsitatea (EHU), Plaza Casilla 3, 48012, Bilbao, Spain.

*Abstract*

A lumped parameter, state space model for the tokamak transformer including the slow flux penetration in the plasma (skin effect transformer model) is presented. The model does not require detailed or explicit information about plasma profiles or geometry. Instead, this information is lumped in system variables, parameters and inputs. The model has an exact mathematical structure built from energy and flux conservation theorems, predicting the evolution and non linear interaction of the plasma current and internal inductance as functions of the primary coil currents, plasma resistance, non-inductive current drive and the loop voltage at an specific location inside the plasma (equilibrium loop voltage). Loop voltage profile in the plasma is substituted by a three-point discretization, and ordinary differential equations are used to predict the equilibrium loop voltage as function of the boundary and resistive loop voltages. This provides a model for equilibrium loop voltage evolution, which is reminiscent of the skin effect. The order and parameters of this differential equation are determined empirically using system identification techniques. Fast plasma current modulation experiments with Random Binary Signals (RBS) have been conducted in the TCV tokamak to generate the required data for the analysis. Plasma current was modulated in Ohmic conditions between 200kA and 300kA with 30ms rise time, several times faster than its time constant $L/R \approx 200ms$. A second order linear differential equation for the equilibrium loop voltage is sufficient to describe the plasma current and internal inductance modulation with 70% and 38% fit parameters respectively. The model explains the most salient features of the plasma current transients, such as the inverse correlation between plasma current ramp rates and internal inductance changes, without requiring detailed or explicit information about resistivity profiles. This proves that lumped parameter modeling approach can be used to predict the time evolution of bulk plasma properties such as plasma inductance or current with reasonable accuracy; at least in Ohmic conditions without external heating and current drive sources.

## I .Introduction

Lumped parameter modelling reduces partial differential equations of a continuous (infinite dimensional) distributed parameter system into ordinary differential equations with a finite number of parameters. Lumped parameter elements arose originally in electronics and electrical engineering, to describe the energy storing and dissipation elements of electric circuits. Generally speaking, lumped parameter modelling can be used to model circuits of short characteristic dimensions compared with the wavelength of operation.

Standard lumped parameter modelling treats a tokamak as a toroidal transformer with one turn secondary R, L plasma ring circuit coupled with a primary transformer circuit, where R,L denote plasma resistance and inductance respectively. The presence of non ohmic current is treated using an additional non inductive current source [1],[2]. The coupling between transformer primary and plasma is accounted by a mutual inductance M. This standard description, despite its simplicity, has proved to be very useful in analysis and prediction of plasma discharge evolution. It has been used for years to infer non inductive current drive fractions and efficiencies in tokamaks [3],[4] and also to perform initial estimations of central solenoid magnetic flux in tokamak reactor studies [5] . However, it does not take into account the variations of plasma inductance that occur during plasma current transients due to the slow flux penetration inside the plasma, or skin effect. This is an important issue for tokamaks, particularly during the initiation and termination phases of the discharge, due to the large machine size and plasma temperature. We have developed an improved transformer model that includes a lumped parameter formulation for the skin effect, and whose purpose is to predict with reasonable accuracy the inductance and current changes as function of the external PF currents, plasma resistance and current drive. We call this model a skin effect transformer model to differentiate it from the standard transformer model where plasma inductance is a fixed parameter.

Distributed parameter simulations are the preferred option to simulate current profile evolution and the associated internal inductance changes that occur during plasma current transients [6] - [13]. These distributed parameter simulations predict strong correlations between internal inductance and plasma current transients, in agreement with the experiments. A correlation model is some kind of lumped parameter formulation, in the sense that does not contain explicit information about profiles. Since plasma current can be predicted with a lumped parameter model, the correlation observed between internal inductance and plasma current transients is evidence suggesting that a lumped parameter formulation could predict internal inductance behaviour. Previous lumped parameter modelling of the JET tokamak has proved the feasibility of this approach [14] .

To be able to describe the effect that plasma current profile evolution has on the internal inductance using the lumped parameter approach, some profile and geometric information must necessarily be embedded (lumped) in the model inputs, parameters and state variables. But the aim the lumped parameter modelling philosophy is not to describe plasma profiles or geometry evolution in itself. For this, distributed parameter models and simulations are required. There are plenty of such distributed parameter models available [15] - [21] , some of them are control oriented [22] , [23] and some of them are even available in real time [24] . From this point of view, lumped parameter models are no substitute for distributed parameter models. However, when it comes to control systems design, a lumped parameter formulation is preferred. Lumped parameter models have been the first modelling choice for tokamak magnetic control for years [25]-[30] , since control system's design methodology is well developed for lumped parameter systems [31] - [34]. The main motivation for the skin effect transformer model is to be used in the design of tokamak plasma current, boundary flux and inductance control systems. An ongoing project at TCV aims to test this type of controls. However Tokamak magnetic control is not the subject this paper. The reader is referred to [35] , [36] and references therein for details of how plasma current and general magnetic control systems are designed. The motivation, application and demonstration of inductance control are a reported elsewhere in the literature [37]-[42] and not discussed here either.

The paper is organised as follows:

Section **II** starts the lumped parameter modelling by identifying some energy storing and dissipation elements of the tokamak circuit, along with some useful relationships among them. A set of exact coupled differential equations for internal inductance and plasma current are derived from energy conservation and power balance theorems. An additional boundary flux conservation theorem provides the link between external coil currents, plasma current and boundary flux. The result is a novel description of the tokamak transformer using only lumped parameters, predicting the evolution and non linear interaction of the plasma current and internal inductance as functions of the primary coil currents, plasma resistance, non-inductive current drive and an additional *equilibrium state* derived from the flux linked by the plasma current distribution. This lumped parameter formulation does not require information about the plasma profiles spatial distribution. Instead, all the information about the profiles is embedded in the state variables and inputs.

Section **III** introduces a simplified model relating the *equilibrium state* to the boundary and resistive voltages. This is required to close the system of equations obtained in section **II**. The equilibrium state model is a lumped parameter version of the skin effect obtained by replacing the loop voltage profile eigenfunction by a three-dimensional eigenvector.

Section **IV** writes the equilibrium state model obtained in section **III** as a generic family of state space models, which is the preferred structure for the analysis of sections **VII** and **VIII**.

Section **V** explains the set-up and details of dedicated plasma current modulation experiments performed at the TCV tokamak. To expose the model non-linearities, the experiments were conducted using an extended range of operation both in amplitude and frequency. This differs from the conventional approach in system identification, where small signal models are sought. Some details of the TCV current control system are explained since they are central to the discussion, but this control system is part of earlier developments at TCV, and is not a contribution of this work.

Section **VI** explains how the data required for the analysis is generated with an equilibrium code, and some basic spectral analysis of the resulting inputs and outputs to give an idea of the experiment's modulation bandwidth.

Section **VII** performs a linear system identification analysis to find empirically the equilibrium state model order and parameters from the experimental data.

Section **VIII** performs a non linear system identification analysis using the model parameters found in section **VII** as an initial guess. A final cross-correlation study between the model residuals and its inputs is used to illustrate the accuracy of the results.

Finally, section **IX** presents a reduced version of the model written in control affine form, so it can readily be used for control systems design.

### II. Derivation of exact tokamak transformer lumped parameter equations

We are after a lumped parameter formulation for the tokamak transformer not requiring explicit information about plasma profiles spatial distribution, but still being able to describe processes related to plasma profiles spatial evolution, such as the skin effect. Necessarily, some kind of profile information must be lumped in some states and inputs. This section identifies state variables and inputs that lump the profile information along with basic relationship among them. These relationships are derived from conservation laws, so the resulting model is exact.

In the following, both a cylindrical coordinate system $(r, \phi, z)$ and a flux coordinate system $(\rho, \theta, \phi)$ are used, and the plasma is assumed to be axisymmetric about the z-axis. Only the time evolving components $(B_r, B_z)$ of the *poloidal* magnetic field and the toroidal components of the electric field $E$ are considered in the analysis.

The region of integration will be defined as the region where there is plasma. This will correspond to a plasma volume $G$, or a plasma cross section $\Omega$, delimited by the plasma boundary $\Gamma$.
The plasma current is defined as

$$I = \int_\Omega j \, dS \qquad (1)$$

where $dS = dr dz$ and $j$ is the toroidal current density.
The poloidal flux function $\psi(r,z)$ is the flux through an arbitrary circle of radius r centred at the torus symmetry axis at a position z. All the points with equal flux define a flux surface. The flux surface surrounding the plasma region is the boundary flux surface $\psi_B$.

The Green´s function for a unit toroidal current ring at $\mathbf{r}_j = (r_j, z_j)$ satisfies
$$\Delta^* G[\mathbf{r}, \mathbf{r}_j] = 2\pi \mu_0 r \delta(r - r_1) \delta(z - z_1)$$
where $\Delta^*$ is the elliptic operator

$$\Delta^* = \frac{1}{r}\frac{\partial}{\partial r}\left(\frac{1}{r}\frac{\partial}{\partial r}\right) + \frac{\partial^2}{\partial z^2} \qquad (2)$$

The mutual inductances $M_j$ between the coil system j and the plasma boundary can be written in terms of Green's functions as [44], [45]:

$$M_j = \mu_0 \frac{1}{2\pi} \int_0^{2\pi} G[\mathbf{r}(a), \mathbf{r}_j] d\theta \qquad (3)$$

Similarly, the external inductance between the plasma current and the plasma boundary is

$$L_e = -\mu_0 \int_0^{2\pi} \chi(a, \theta_1) G[\mathbf{r}(a), \mathbf{r}_1(a)] dl(\theta_1) \qquad (4)$$

where $dl(\theta)$ denotes the differential arclength along the coordinate $\theta$ and

$$\chi(\rho, \theta) = \frac{|\nabla \rho|}{r \int_0^{2\pi} \frac{|\nabla \rho|}{r} dl(\theta)} \qquad (5)$$

Practical numerical details on how to calculate Green's functions and avoid the singularity at $\mathbf{r} = \mathbf{r}_1$ can be found in [44], [45]

Using the external and mutual inductances, the flux at the plasma boundary $\psi_B$ can be written as the sum of the contributions from the PF coil systems and the internal plasma current distribution, according to the boundary flux conservation equation:

$$\psi_b = L_e I + \sum M_j I_j \tag{6}$$

The plasma internal inductance $L_i$ is defined from the magnetic energy content inside the plasma volume $G$

$$W = \frac{1}{2} L_i I^2 = \frac{1}{2\mu_0} \int_G \left( B_r^2 + B_z^2 \right) dv \tag{7}$$

where $\mu_0$ is the vacuum magnetic permeability and the differential volume element is

$$dv = r dr d\phi dz \tag{8}$$

The quantity W contains magnetic energy created by the plasma current as well as by external conductors.
The plasma inductance $L_p$ is the sum of the internal and external inductance components.

$$L_p = L_e + L_i \tag{9}$$

An alternative expression for the magnetic energy (7) as a function of poloidal fluxes at certain locations and the current density is [14]

$$W = \frac{(\psi_C - \psi_B) I}{2} \tag{10}$$

and equivalently

$$L_i I = (\psi_C - \psi_B), \tag{11}$$

where $\psi_C$ is defined as

$$\psi_C = \frac{\int_\Omega \psi j dS}{I} . \tag{12}$$

The flux $\psi_C$ is termed *equilibrium flux*, as it is the sole variable in our treatment that depends explicitly on the details of the plasma equilibrium. Although $\psi_C$ is a distribution average, we can formally identify a flux surface inside the plasma on which the flux takes the value $\psi_C$. This flux surface, which may of course move radially in time, will be termed the *equilibrium flux surface*.

The Poynting energy theorem in toroidal geometry is written as *[1]*

$$\frac{dW}{dt} + \int_\Omega jV dS = V_B I \tag{13}$$

or equivalently as

$$\frac{dW}{dt} = (V_B - V_R)I \tag{14}$$

where the loop voltage at the boundary flux surface is

$$V_B = -\frac{d\psi_B}{dt} \tag{15}$$

and the resistive drop $V_R$ is written in terms of the total plasma current, resistance and an equivalent non-inductive current $\hat{I}$ as

$$V_R = R(I - \hat{I}) \tag{16}$$

The resistive flux is defined as

$$\psi_R = -\int_0^t V_R \, dt = -\int_0^t R(I - \hat{I}) \, dt \tag{17}$$

Again, we can formally identify a flux surface inside the plasma on which the flux takes the value $\psi_R$. This flux surface, which may of course move radially in time, will be termed the *resistive flux surface*.

The ohmic power dissipation from Poynting analysis is given as a volume integral of the current density distribution $j$, effective plasma resistivity $\eta$, and non-inductive current density $\hat{j}$ in the toroidal direction.

$$\int_\Omega Ej \, dv = \int_\Omega \eta(j - \hat{j})j \, dv = \int_\Omega \eta j^2 \, dv - \int_\Omega j\eta\hat{j} \, dv \tag{18}$$

The ohmic power dissipation in the lumped parameter picture is

$$R(I - \hat{I})I = RI^2 - RI\hat{I} \tag{19}$$

both descriptions are desired to be the same, so power balance is fulfilled:

$$RI^2 - RI\hat{I} = \int_\Omega \eta j^2 \, dv - \int_\Omega j\eta\hat{j} \, dv \tag{20}$$

so the natural choices are

$$RI^2 = \int_\Omega \eta j^2 \, dv \tag{21}$$

$$RI\hat{I} = \int_\Omega j\eta\hat{j} \, dv \tag{22}$$

leading to the following expressions for plasma resistance $R$ and current drive $\hat{I}$:

$$R = \frac{\int_\Omega \eta j^2 \, dv}{I^2} \tag{23}$$

$$\frac{\hat{I}}{I} = \frac{\int_\Omega j\eta \hat{j}\, dv}{\int_\Omega \eta j^2\, dv} \tag{24}$$

These are the only definitions that make compatible the Poynting power balance analysis (13) with (14)

Linear combinations of (14) and of the time derivative of (11) lead to the following set of coupled equations

$$I\frac{dL_i}{dt} = 2(V_R - V_C) \tag{25}$$

$$L_i \frac{dI}{dt} = V_B + V_C - 2V_R \tag{26}$$

The equilibrium voltage at the equilibrium flux surface is derived from Lenz's law

$$V_C = -\frac{d\psi_C}{dt} \tag{27}$$

Equations (25) and (26) determine the evolution of the plasma current and internal inductance as functions of boundary size and shape, equilibrium, and resistive voltages.

The final steady-state solution for plasma current and inductance corresponds to a constant loop voltage profile across the plasma:

$$\frac{\partial \psi(r,t)}{\partial t} = V_B = V_R = V_C. \tag{28}$$

The flux diffusion dynamics is embedded in this model. For instance, a drop in plasma resistance at constant boundary voltage will increase the plasma current, which in turn will increase the resistive drop (16). Similarly, an increase in boundary loop voltage will drive more plasma current and the resistive drop will also increase.
Using the flux balance equation (6) and Ohm's law (16), the dynamic response (25), (26) can be written explicitly as a function of PF coil currents, plasma resistance and non inductive current drive:

$$I\frac{dL_P}{dt} = 2R(I - \hat{I}) - 2V_C + I\frac{dL_e}{dt} \tag{29}$$

$$L_P \frac{dI}{dt} = V_C - 2R(I - \hat{I}) - \sum M_j \frac{dI_j}{dt} - I\frac{dL_e}{dt} - \sum I_j \frac{dM_j}{dt} \tag{30}$$

Combination of (6), (15), (29), (30) leads to an expression for the boundary voltage as direct function of the states, inputs and model parameters:

$$V_B = -\frac{L_e}{L_P}\left(V_C - 2R\left(I - \hat{I}\right)\right) - \frac{L_i}{L_P}\sum M_j \frac{dI_j}{dt} - \frac{L_i}{L_P}\sum I_j \frac{dM_j}{dt} \tag{31}$$

The differential equations (29),(30) (or alternatively (25), (26),(31)) predict the exact evolution of plasma current and inductance without requiring explicit information about plasma profiles spatial distribution. Instead, all the information about the profiles is lumped in the state variables, parameters and inputs. Note that (29),(30) is not yet a closed set of equations. Equations (29),(30) depend on the equilibrium state (27), whose dynamics will determined empirically in later sections using system identification techniques.

The plasma is kept within the vacuum vessel boundaries by means of additional plasma shape and position control systems acting on the PF coils. When shape and position are kept approximately constant, the terms on the right hand side containing time derivatives of mutual and external inductances can be neglected, and the set of equations simplifies to

$$I \frac{dL_P}{dt} = 2R\left(I - \hat{I}\right) - 2V_C \tag{32}$$

$$L_P \frac{dI}{dt} = V_C - 2R\left(I - \hat{I}\right) - \sum M_j \frac{dI_j}{dt} \tag{33}$$

If in addition the plasma internal inductance is constant in time, a standard transformer equation is obtained

$$-\sum M_j \frac{dI_j}{dt} = R\left(I - \hat{I}\right) + L_P \frac{dI}{dt} \tag{34}$$

The physical meaning of this equation is that the change of flux produced by the external coils generates a loop voltage that compensates the resistive drop and builds up the plasma current. According to (34), the time constant for inductive current build up is

$$\tau_p = L_p / R \tag{35}$$

with $R$ and $L_p$ given by (23) and (9).

**III Lumped parameter model approximation for the skin effect**

Equations (29),(30) are exact. They predict the evolution and non-linear interaction of the plasma current and internal inductance as functions of the primary coil currents, plasma resistance, non-inductive current drive and equilibrium loop voltage. To obtain a closed set of equations, we still have to obtain a model for the equilibrium voltage (27) state dynamics as function of the remaining system states and inputs of the model.

We want to find a lumped parameter version of the flux diffusion equation as function of the available states and inputs alone. The required output of the model is the equilibrium state. To build the lumped parameter equivalent to flux diffusion equation we have to determine what system states and inputs are relevant. To do this, we will replace the loop voltage profile eigenfunction by a three-dimensional eigenvector.

At any plasma location, the current density and vector potential are related by Poisson´s equation

$$\mu_0 \mathbf{j} = -\nabla^2 \mathbf{A} \qquad (36)$$

Writing the vector potential as a function of poloidal flux, Poisson´s equation can be written for the toroidal component of the current density as

$$2\pi\mu_0 r j = -\Delta^* \psi \qquad (37)$$

where $\Delta^*$ is the elliptic operator (2) .

Using Ohm's law, the voltage at any plasma location, and in particular $V_C$, is a solution of

$$V = -\frac{\eta}{\mu_0}\Delta^*\psi + 2\pi\eta r \hat{j} \qquad (38)$$

Taking the time derivative of this expression leads to

$$\frac{\partial V}{\partial t} = \frac{1}{\mu_0}\left(-\frac{\partial \eta}{\partial t}\Delta^*\psi + \eta\Delta^* V\right) + 2\pi\frac{\partial}{\partial t}\left(\eta r \hat{j}\right) \qquad (39)$$

This distributed parameter equation is exact, but since the flux, resistivity and spatial distribution of non inductive current density are assumed to be unknown, it can not be directly incorporated in the model.

The simplest form of (39) corresponds to fixed plasma geometry with time invariant resistivity and non inductive current profiles;

$$\frac{\partial V}{\partial t} = \frac{\eta}{\mu_0}\Delta^* V \qquad (40)$$

Thus, under these conditions, the time derivative of the loop voltage is proportional to the second spatial derivative of the loop voltage profile.

If $\rho_R < \rho_C < \rho_B$ are the effective radii of the flux surfaces $\psi_R, \psi_C, \psi_B$, and the loop voltage profile varies smoothly across the plasma, we can approximate the second order spatial derivative by finite differences at neighbouring points around $\rho_C$

$$\frac{\partial V_C}{\partial t} \propto \frac{(V_B - V_C)}{\Delta\rho_{BC}} - \frac{(V_C - V_R)}{\Delta\rho_{CR}} \propto -\frac{(\Delta\rho_{BR})}{\Delta\rho_{BC}\Delta\rho_{CR}}V_C + \frac{V_B}{\Delta\rho_{BC}} + \frac{V_R}{\Delta\rho_{CR}} \qquad (41)$$

where

$$\Delta\rho_{BC} = \rho_B - \rho_C$$
$$\Delta\rho_{CR} = \rho_C - \rho_R \qquad (42)$$
$$\Delta\rho_{BR} = \rho_B - \rho_R$$

Equation (41) has already the shape of a stable first order differential equation for $V_C$ in which the resistive drop and the boundary voltage add up as a compound input to the system. It also has the required steady state solution (28). The metric (42) is unknown, and is also a function of time, as the equilibrium evolves. However, despite the numerous unknowns, (41) provides a useful insight into what states and inputs could be used in a linear state space model. This equation suggests in particular a state space model relating the equilibrium voltage $V_C$ to a linear combination of $V_R, V_B$. We can also infer that, since the metric is expected to evolve in time, a first order structure with fixed coefficients is unlikely to be an accurate description, but we will not exclude this possibility from the start.

**IV State space model for equilibrium voltage**

For system identification purposes, we find convenient to write the dynamics of the equilibrium voltage (41) as a generic family of state space models of order *n* written in innovations form:

$$\frac{d\mathbf{x}}{dt} = \mathbf{A}\mathbf{x} + \mathbf{B}\mathbf{u} + \mathbf{K}r \qquad (43)$$
$$\mathbf{y} = \mathbf{C}\mathbf{x} + r$$

Where x is a state space vector, $y = V_C$ is the output equation and

$$u = \begin{pmatrix} V_B & V_R \end{pmatrix}^T \qquad (44)$$

is the input vector. The residuals *r* will represent all features present in the output that are not explained by the model. A steady state Kalman [46] matrix **K** is incorporated to reconstruct frequency response diagrams, including the noise power spectrum, perform residuals selfcorrelation (whiteness tests) and input-residuals correlations (independence tests). This will discussed in detail in later sections.

The model order *n* and the matrices $\mathbf{A}, \mathbf{B}, \mathbf{K}$ can then be determined from experimental data using system identification techniques. For these techniques to work we need to dynamically stimulate the plasma to obtain loop voltage data with broad spectral content. To generate the required data for the analysis closed loop experiments using random binary signals (RBS) imposed as plasma current references have been conducted in the TCV tokamak. The preparation and details of these experiments are discussed in the next section

**V System identification experiments**

The 'Tokamak à Configuration Variable' TCV is a fusion plasma research facility particularly suited for the study of advanced plasma control systems [47]. It has a major radius of 0.9m, a typical plasma volume of ~1.5 m$^3$, and a great flexibility of plasma shape configurations.
TCV has an air core transformer with an available flux swing of 3.4 Webers, Ohmic power input of about 1MW and 4.5 MW of additional Electron Cyclotron Resonance Heating (ECRH) power. This results in typical plasma discharge duration of 2 seconds and a maximum plasma current up to 1.0 MA, with loop voltage transients up to 10V.

On TCV, the plasma current is made to follow a prescribed waveform by acting on the primary transformer coil voltage. Plasma current feedback is necessary to compensate for the changes in plasma resistance and non inductive current. For a detailed description of TCV hybrid control system the reader is referred to [48]

Ideally, plasma current feedback control ought to be disabled to test the open loop response of the system. This is however not practical, since without current feedback the plasma current undergoes uncontrolled excursions, leading to plasma disruptions in many occasions. The situation is particularly acute in the extreme modulation conditions of the experiment discussed later in this section. We are then forced to maintain the plasma current feedback and adopt a closed loop system identification approach.

Feedback systems introduce correlations between the perturbations in the feedback loop and the inputs. Unmodeled perturbations are treated as noise. When inputs and noise are comparable (as is the case when small signal models are being sought) closed loop system identification is better performed by adding a known perturbation inside the feedback loop [[48] ],[49]. However, the high gain in the TCV current feedback loop implies that any perturbations introduced inside the loop (e.g. adding a feed-forward voltage at the ohmic coil, as in [49]) are quickly attenuated, so the perturbation technique was not practically compatible with the need to limit the current transients.

Closed loop system identification, however, can be successfully performed when the system is stimulated with large signals compared with the noise input, and/or a good model for the noise input exists, so noise becomes a modeled perturbation. We are in fact interested in exploring the system's non-linearity in an extended range of operation both in amplitude and frequency, and the perturbation is largely coming from plasma resistance variations, for which we have a very accurate model at our disposal. This will allow us to use a direct closed loop system identification approach to extract the relevant open loop dynamic information, in both the frequency and time domains.

To excite high order dynamics of the plasma current / inductance over a broad spectrum, RBS waveforms are imposed as references for the plasma current during the flat top. Plasma current was modulated between 200kA and 300kA with 30ms rise time, several times faster than its time constant $L_p/R \approx 200$ms. A feedback control system generates primary transformer voltage control actions to drive the desired plasma current. The plasma current feedback system has a settling time of less that 8ms, negligible compared with the time constant (35) $\tau_p \approx 200$ms . The resulting voltage signals for the primary transformer coil and the corresponding responses in primary coil current and plasma current are shown in figures 1 and 2. The time derivative of the external flux created by the transformer coil and the additional PF coil currents create a plasma voltage that for constant plasma and mutual inductances can be approximated by (34). The primary transformer current excitation and the resulting plasma current are shown in Fig. 1 for two system identification experiments. A positive current in the toroidal direction flows in the direction of increasing toroidal angle. The sign criteria are given by Lenz's and Ohm's law. A positive voltage applied to the transformer primary generates a current that increases in time during the discharge. The corresponding boundary flux that increases in time will generate negative boundary voltage and negative plasma current.

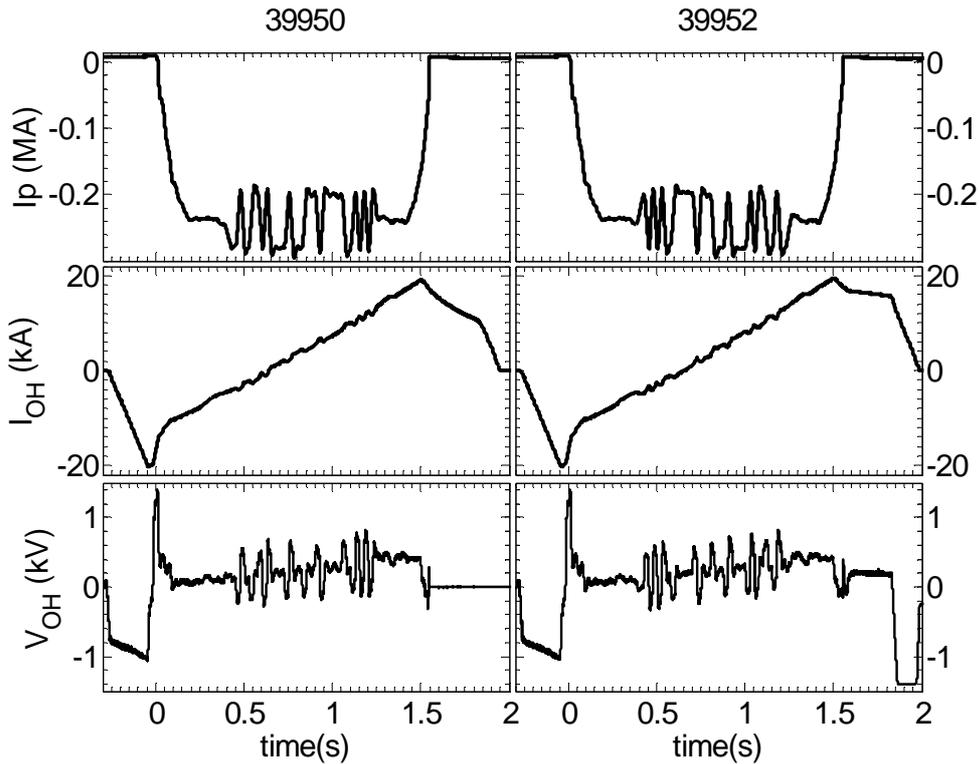

Fig. 1. System identification experiments using Random Binary Signal stimulation for plasma current (top) along with the required current (middle) and power supply voltage (bottom) in the transformer primary. Two system identification experiments are shown. Plasma current is modulated in amplitud several times faster than its natural time constant.

The time prior to the plasma current build up (shown as negative) is used to charge the primary transformer with negative current, to increase the available flux swing. Then, the transformer current is brought up to positive values to drive negative plasma current. The time origin $t_0 = 0$ is chosen as the time when the poloidal magnetic field null necessary for breakdown is created inside the vacuum vessel. Figure 2 shows the same quantities during the RBS time window, with the offsets and linear trends removed.

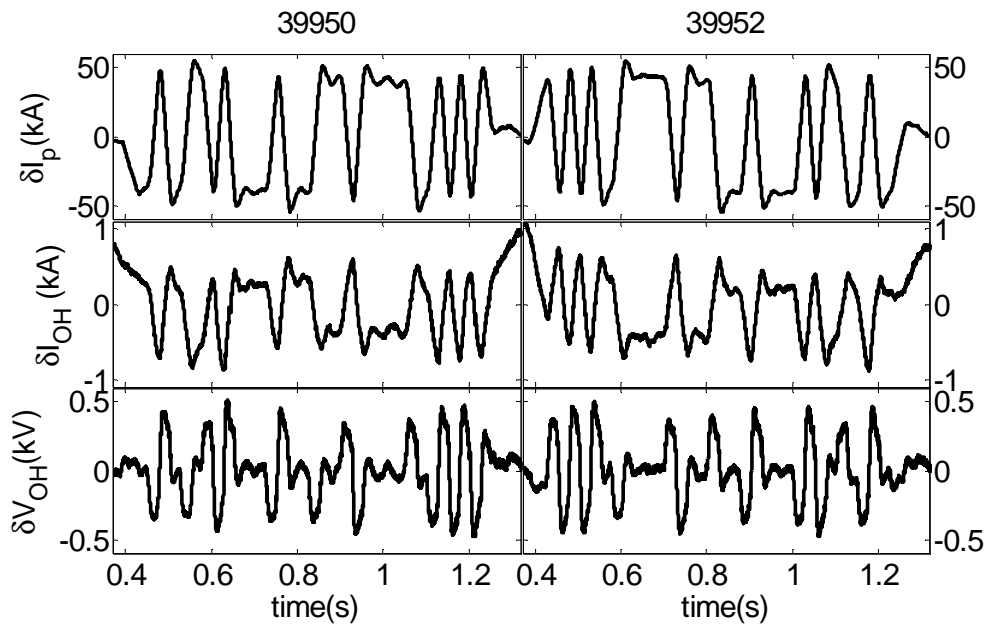

Fig. 2. Plasma current, transformer primary current and voltage from Fig. 1 after removing offsets and linear trends.

## VI Preliminary data generation and analysis

The required inputs and outputs to the system cannot be measured directly. The data used for the validation is obtained from the equilibrium reconstruction code LIUQE [50] . This code determines magnetic flux surface distribution from a set of external magnetic measurements [51] . Among other plasma variables, LIUQE delivers the plasma current, boundary flux and *normalised* internal inductance that are required for the analysis. Plasma current and internal inductance can also be obtained from magnetic probes surrounding the plasma [52] but we prefer to use the LIUQE's plasma current to obtain a consistent data set. Similar results can also be obtained using transport simulations [53] .The plasma internal inductance is defined relative to the machine major radius ( $r_0 = 0.88m$ on TCV) and the *normalised* internal inductance, using

$$L_i = \mu_0 r_0 \frac{l_i}{2} \qquad (45)$$

The resistive voltage is derived from the power balance (14) as

$$V_R = V_B - \frac{1}{2I}\frac{d}{dt}(L_i I^2) \qquad (46)$$

The equilibrium flux is derived from (11) as

$$\psi_C = L_i I + \psi_B \qquad (47)$$

Boundary and equilibrium voltages are obtained from (15) and (27) respectively.

External magnetic measurements from flux loops and probes around the vacuum vessel are acquired at a 2 kHz sampling rate. The reconstructed equilibrium data for the current analysis has been generated with a 1.5 ms spacing.
Figure 3 shows equilibrium reconstructed data for shot 39952 in a short time window at the beginning of the *flat-top*.

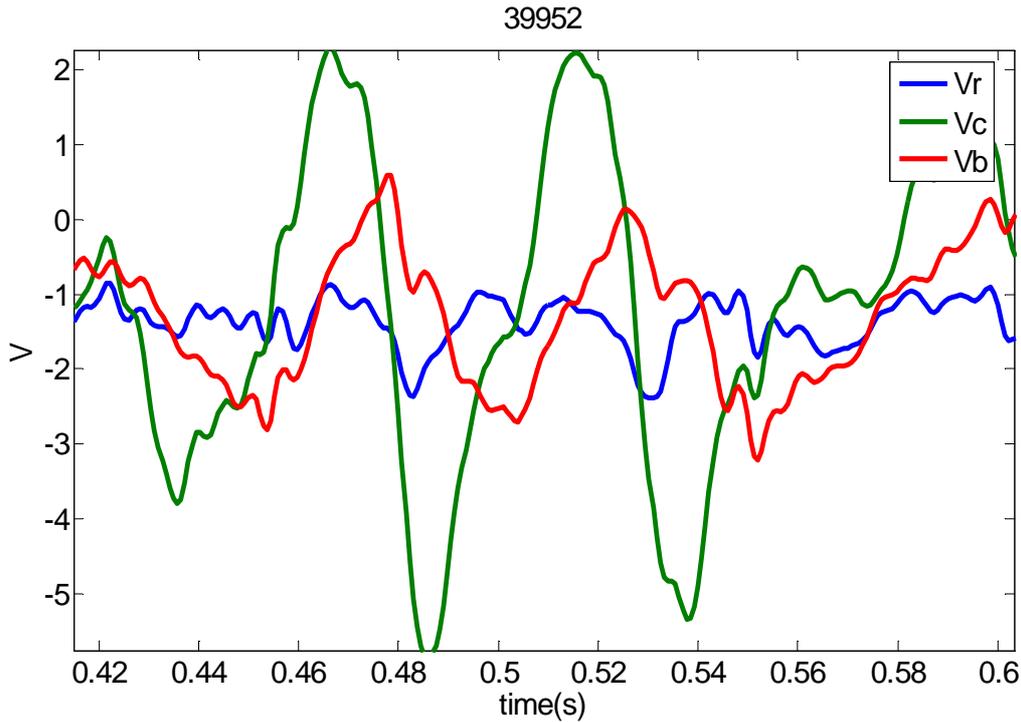

Fig. 3. Evolution of plasma boundary ($V_B$), equilibrium ($V_C$) and resistive ($V_R$) loop voltages for experiment 39952.

The resistive drop $V_R$ input has a poor stimulation compared with the boundary voltage input $V_B$. This makes it difficult to perform a linear system identification analysis using separate inputs $V_B, V_R$ and a single cost function based on $V_C$ alone.

A preliminary linear regression analysis (zero order, no dynamics) correlating the output to a linear combination of the inputs $u = kV_B + (1-k)V_R$ finds a minimum for $k \cong 0.5$, suggesting that a simplified analysis using a single compound input $u = 0.5(V_B + V_R)$ may be a good starting point. The model order will be determined from this in the next section. But first, a preliminary inspection of the reconstructed compound inputs $u = 0.5(V_B + V_R)$ and outputs $V_C$ is worthwhile. This is shown in figure 4. Simple eye inspection of this figure helps to evaluate bandwidth of the system identification experiments. The largest peaks of the voltage oscillations are spaced about 50-100ms, so the main part of the input signal power is carried by low frequency components of less than 20 Hz. The corresponding spectrum estimate is shown in figure 5. The spectral power density of the signals halves above 30 Hz respect to its maximum. Above 70 Hz there is no significant power in inputs or outputs, so it will be very difficult to reconstruct the spectrum (bode diagrams) above this frequency. This will be shown in the next section.

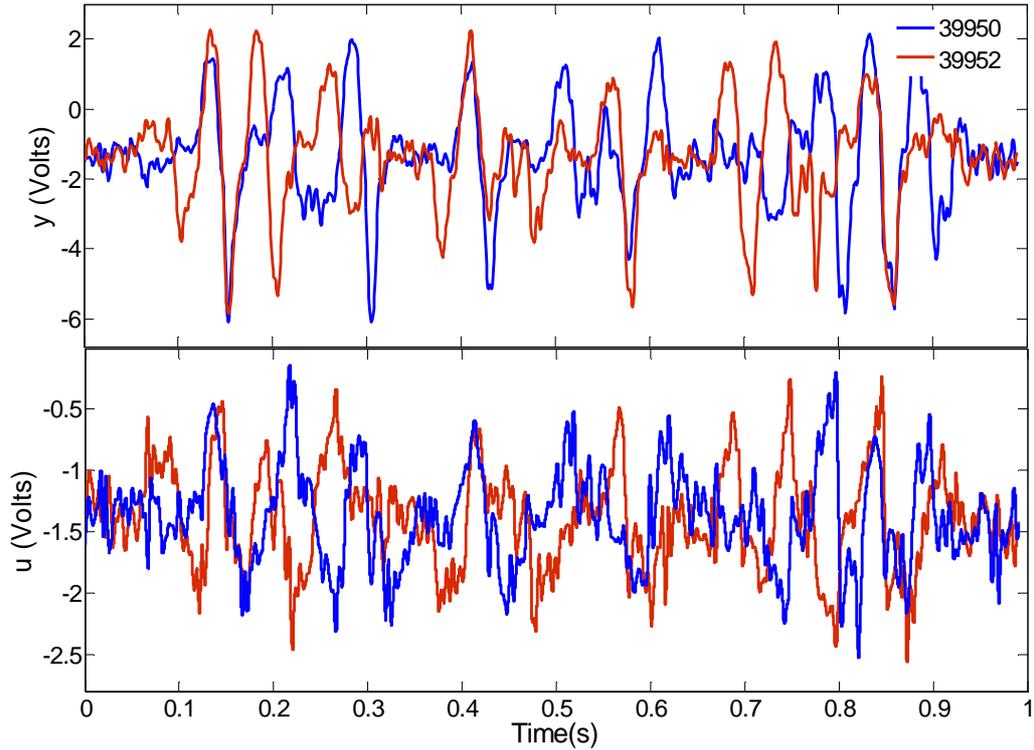

Fig. 4. Input u=0.5($V_B$+$V_R$) and output y=$V_C$ for two system identification experiments.

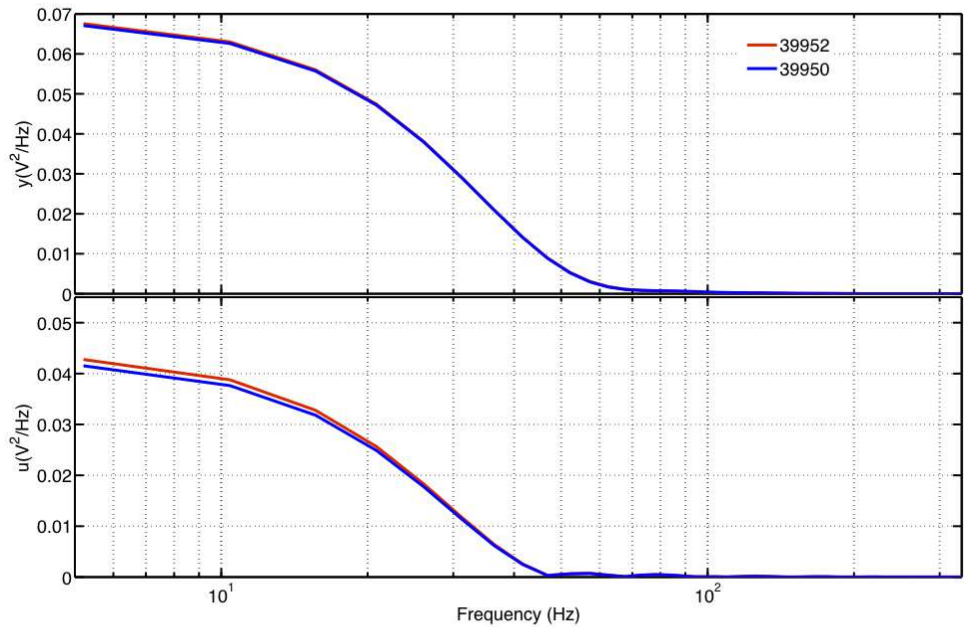

Fig. 5. Power spectra of input u=0.5($V_B$+$V_R$) and output y=$V_C$ signals for two system identification experiments. Both system identification experiments are almost identical in terms of power spectral content. Most of the power is concentrated below 70 Hz.

**VII Linear system identification**

Next, we conduct a linear system identification analysis from actual experimental data, to determine the best order of the model. To do this, the experimental data from a RBS stimulated discharge (39950) is used to fit the model parameters using the prediction error method. The parameters found are then used to predict the behavior of another discharge

(39952) with an infinite prediction horizon (simulation). Finally, the equilibrium voltage simulation is compared with the actual experiment, and a loss function $R_N$ is derived from the residual $\varepsilon$.

$$\varepsilon = (y_s - y_x) \tag{48}$$

$$R_N^2 = \frac{1}{N}\sum_{i=1}^{N}(y_s - y_x)^2 \tag{49}$$

where N is the number of experimental $y_x$ and simulated $y_s$ data points in the prediction set. The sample variance of the data set is defined as

$$S_N^2 = \frac{1}{N}\sum_{i=1}^{N}(y_x - \bar{y}_x)^2 \tag{50}$$

and the overall fit parameter is defined as

$$f = 1 - \left(\frac{R_N^2}{S_N^2}\right)^{\frac{1}{2}} \tag{51}$$

The Akaike Information criteria (AIC) and forward prediction error (FPE) are used to measure the quality of the prediction and best system order [54]. AIC and FPE not only reward the goodness of fit, but also include a penalty that is an increasing function of the number of estimated parameters. According to Akaike's theory, the most accurate model has the smallest AIC and FPE.

The AIC and FPE are defined by the following equations:

$$AIC = \log R_N^2 + \frac{2d}{N} + C \tag{52}$$

$$FPE = R_N^2 \left(\frac{1 + \frac{d}{N}}{1 - \frac{d}{N}}\right) \tag{53}$$

where d is the number of estimated parameters, and C is a constant that that be ignored in model comparisons [54].

Figure 7 shows the simulation results for models up to seventh order.

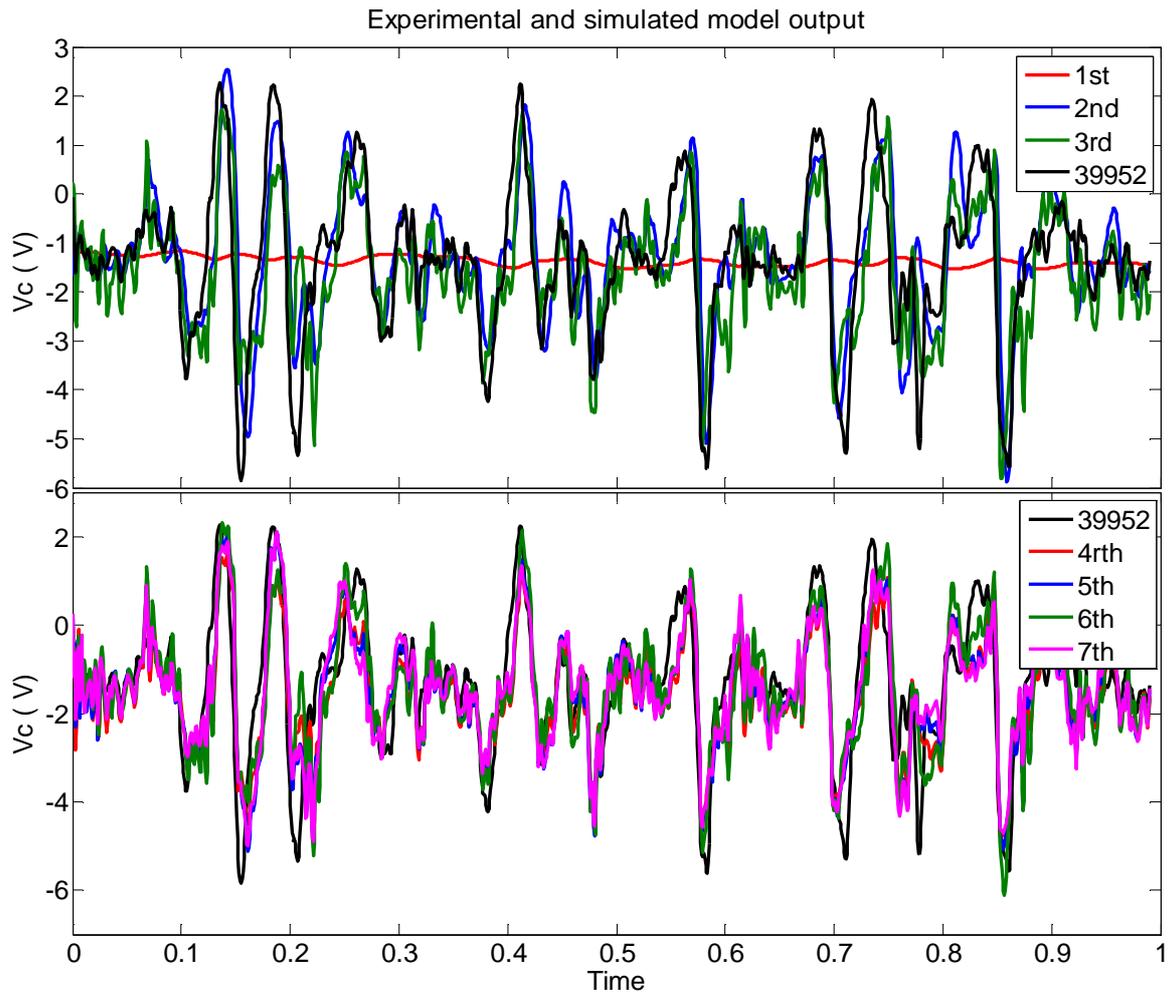

Fig. 6. Experimental data for RBS stimuli on TCV and simulated model output for linear models up to 7$^{th}$ order.

The fit parameter for the first order model (red trace) is 1.5%. Clearly, a first order system (red trace) is not sufficient to describe the dynamics. This supports our initial insight, stemming from the observation that the metric of (41) evolves in time along with the loop voltage profile.

Generally speaking, the higher the order the lower the loss function will be, at the expense of introducing a larger number of parameters in the model. Despite the good visual agreement of most models of fig. 6, the fit parameter (51) for the second order system simulation is just 25%. This is explained by the small scale noise content features in the identification data set, which are difficult to appreciate in the figure.

Figure 7 shows a comparison of models up to seventh order. An exponential function of the AIC and a square root of the FPE are plotted so the result expressed in volts can be directly compared with the loss function. The AIC and FPE criteria are dominated by the loss function having a minimum for a second order system. The loss function for this case is about 50 mV

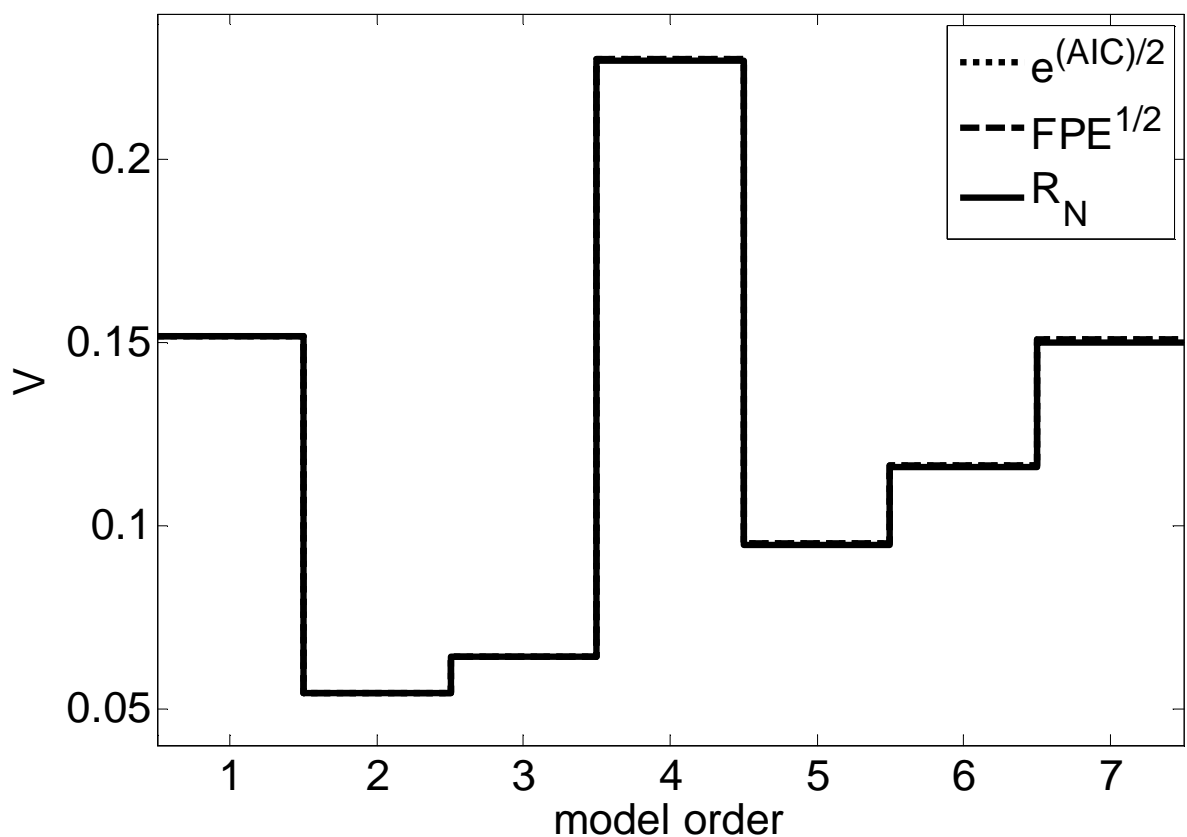

Fig. 7. Comparison of models up to seventh order. An exponential function of the AIC and a square root of the FPE are plotted so the result expressed in volts can be directly compared with the loss function. The AIC and FPE criteria are dominated by the loss function $R_N$ having a minimum for a second order system. The loss function for this case is about 60 mV.

Next we evaluate the impulse response of the system using a non-parametric model [54] derived from the experimental data using correlation analysis. Because nonparametric and parametric models are derived using different algorithms, agreement between these models increases confidence in the parametric model results. The corresponding transient plots provide also valuable insight into the characteristics of model dynamics, such as natural resonances or settling time. We consider the simplest, discrete-time model to estimate the dependence of current and past output values subject to noise in the system. We estimate a high order (n=1 , m=70 ) autoregressive exogenous (ARX) model [54] ,[55],[56] with the structure:

$$a_0 y(t) + a_1 y(t-T) + ... + a_n y(t-nT) = b_0 y(t) + b_1 y(t-T) + ... + b_m y(t-mT) + e(t) \qquad (54)$$

A small negative lag is introduced in the correlation to investigate any residual feedback effects in the input-output data. Then we use this high order model to simulate the system's impulse response. Finally, the parametric models up to seventh order are compared with the correlation model response. Figures 8 and 9 show the results. The first order model is the worst performer. The second order approximation seems closer to the under-damped behavior and natural frequency (about 30Hz) shown by the correlation model.

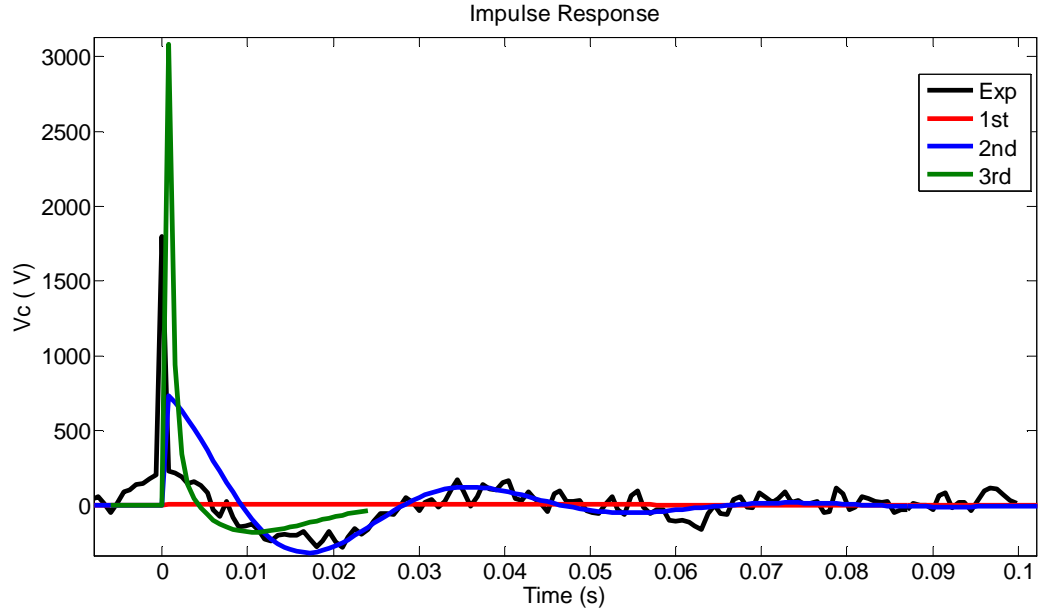

Fig. 8. Impulse response from correlation model (black) and parametric models up to third order (in color)

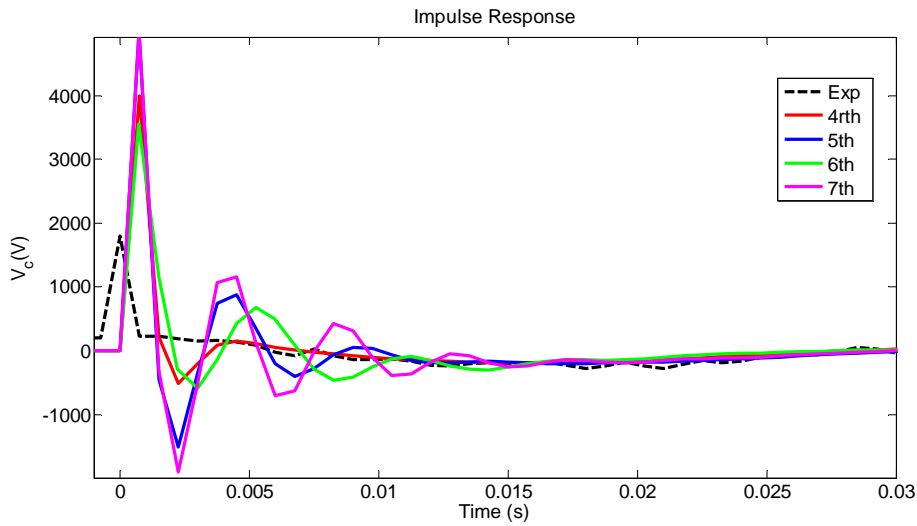

Fig. 9. Impulse response from correlation model (black) and parametric models from 4$^{th}$ to 7$^{th}$ order (in colour)

The next step is to check the whiteness of the residuals and the residual- input dependencies. To do this we use two correlation functions for the model output r and an input u, defined as

$$R_{r,r}(\tau) = \frac{\int_0^{T_{corr}} r(t)r(t+\tau)dt}{\int_0^{T_{corr}} r^2(t)dt} \qquad (55)$$

$$R_{r,u}(\tau) = \frac{\int_0^{T_{corr}} r(t)u(t+\tau)dt}{\left(\int_0^{T_{corr}} r^2(t)dt \int_0^{T_{corr}} u^2(t)dt\right)^{\frac{1}{2}}} \qquad (56)$$

Good models have the self correlation of the residuals (whiteness test) and the cross correlation of the residuals with the input (independence tests) below certain confidence intervals that are calculated from the estimated uncertainty in the model parameters, assuming a Gaussian distribution for the estimates. The results of this analysis are shown in Fig. 10 for models up to order 3. The correlation windows are extended over one characteristic response time of the system (35), the order of 200 ms. The first order system is the worst performer in this set. Second, third and higher orders (not shown) pass the whiteness test, but only the third order passes the correlation test, followed marginally by the second order model.

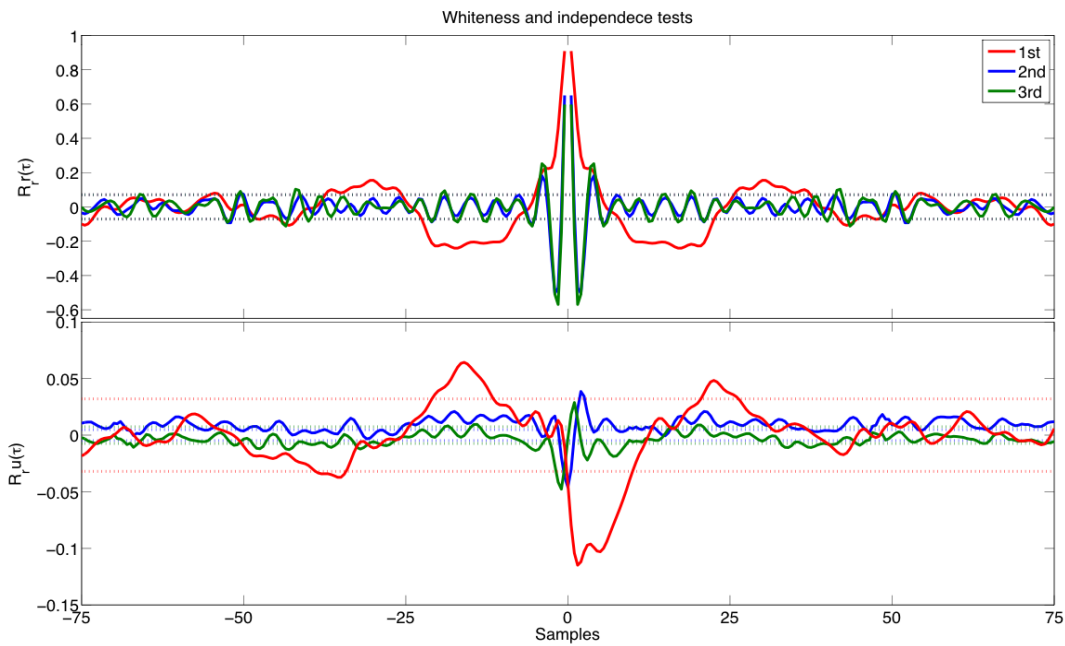

Fig. 10. Whiteness and independence tests for models up to third order. Self correlation of residuals (top) and cross correlation between input and residuals (bottom) are shown for up to 75 lags (112.5 ms). The discontinuous lines bound the range that has a 99% probability of being statistically insignificant.

Figure 11 shows a comparison between the model and experimental data frequency responses. The experimental frequency response is not well determined above 50Hz, particularly in phase, due to the low signal to noise ratio at these frequencies. This can be appreciated in figure 5, where it can be observed than the spectral power density of the input signals halves above 30 Hz respect to its maximum. The first order system response is completely inadequate, whereas the second order response seems again to be closer to experimental results, exhibiting a resonance frequency around 30 Hz. The second order system with two poles and one zero is consistent with the experimental data, despite the noise present. The 45°/decade rise of the zero is arrested by the overlapping of the two poles, each contributing with a 45°/decade drop in the limited range of frequencies where all are active contributors to the phase.

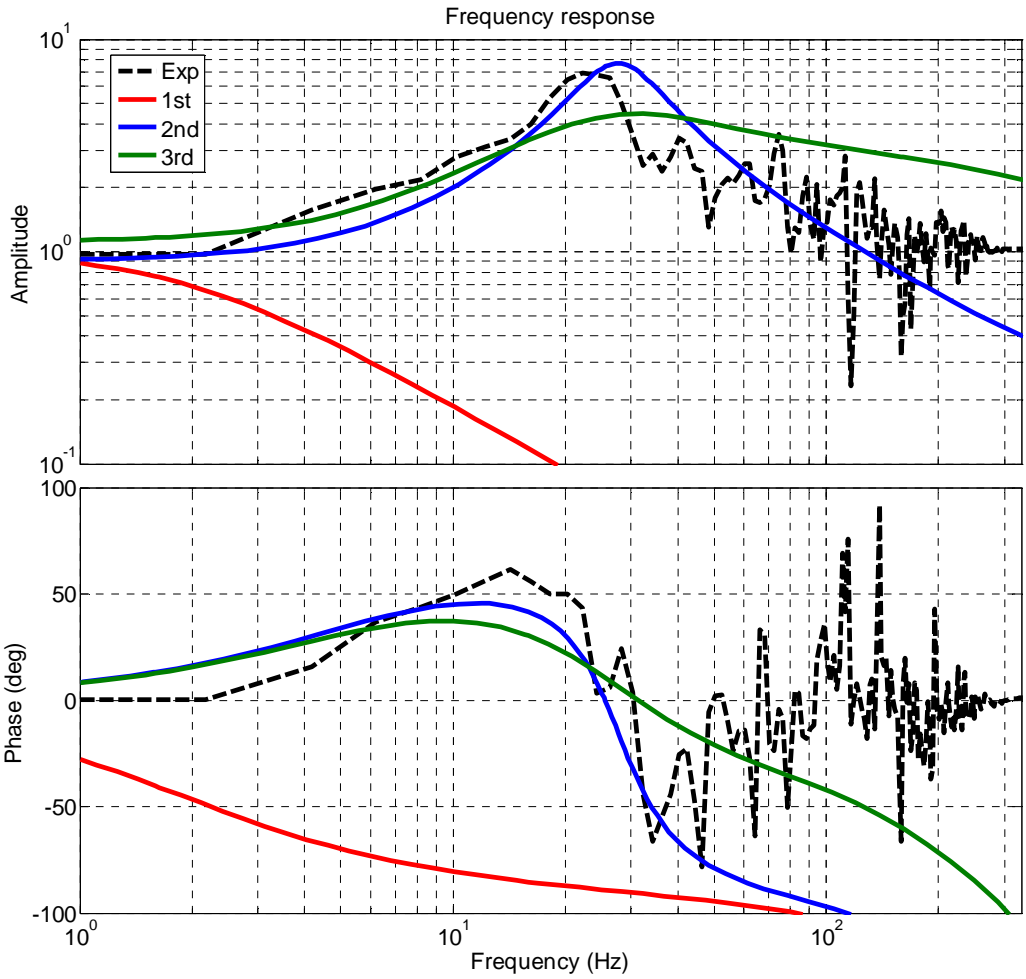

Fig. 11. Amplitude and phase Bode diagram for experimental data (black) and models up to third order (colour) up to the Nyquist frequency of data (333 Hz).

According to all the tests, a second order linear model seems to be the best approximation for equilibrium loop voltage dynamics.

The resulting second order state space model can be described by the following transfer function:

$$G(s) = \frac{y(s)}{u(s)} = \frac{k\omega^2 (1+Ts)}{s^2 + 2\delta\omega s + \omega^2} \tag{57}$$

The parameters found for the second order system using the previous analysis are summarised in table 1

| $\omega$ | $\delta$ | $T$ | $k$ |
|---|---|---|---|
| 174 | 0.3 | 0.012 | 0.9 |

Table 1 Linear system identification of a linear second order model using the equilibrium voltage $V_C$ in the objective cost function.

We generally trust the model order found by the system identification, and believe the frequency $\omega$ at which the equilibrium loop voltage will resonate when driven by an external oscillation (see table 1) is probably in the right magnitude order (tenths of Hz). The steady state gain found (k=0.9) is also quite close to what we expect in steady state (k=1). But we don't trust the damping factor found, for the following physical reasons.

The damping factor is $\delta = 0.3$, allowing resonant natural oscillations on the equilibrium loop voltage even when the system is not driven by an external source. This resonant behaviour requires at least two forms of energy accumulation in the system that can mutually be exchanged. In an electric circuit these are the energy stored in the electric fields (e.g capacitors), and the energy stored in magnetic field (e.g. inductors). In our problem the energy stored in the poloidal magnetic field is about 20 KJ, and the energy stored in the toroidal electric field (<1V/m) is negligible compared with it ($<10^{-11}$ J). The time variations of toroidal electric field are associated to the displacement current. At frequencies below the GHz range these can be neglected in the tokamak analysis. Therefore, if the toroidal electric field was responsible for the effect, resonant behaviour should be observed at frequencies above the GHz range, not in the sub-kHz range as we observe. Another possibility is the coupling with the radial electric fields through the return current [57]. In this case the radial electric fields are the order of kV, but again the energy contribution is again too small ($<10^{-8}$J) compared with the magnetic energy. Yet one more form of energy accumulation is kinetic in nature. During the experiments, plasma radial excursions take place as a response to the internal current redistribution. In this experiments this can take place at maximum speeds of 0.5 m/s. This gives a plasma kinetic energy term [28] which is again negligible ($<10^{-7}$ J) compared with the energy stored in the magnetic field, so it could not be responsible for the resonance found. None of the above forms of secondary energy accumulation can explain the resonant behaviour in the sub-kHz range. We should have obtained instead a damping factor larger than one, corresponding with a damped system $\delta \geq 1$ with two real poles (*two time scale dynamics*), which does not allow oscillations unless the system driven by an external source.

We believe the observed resonant behaviour may be due to artefacts introduced by the equilibrium reconstruction, or perhaps due to the fact that we have used a single compound input for the analysis. There are also inputs to the system that have been assumed to be small and have not been modelled in the equilibrium voltage dynamics either, like bootstrap current. There are also conceivable uncertainties in the equilibrium reconstruction that are not considered, besides of course, all features that can not be explained due to the limitations of the lumped parameter approximation. All these factors are considered to be small contributors, and are included in the model as small perturbations or noise, which are accounted by means of the Kalman noise model structure of (43). The noise spectrum of these is presented in figure 12 exhibiting a maximum that suspiciously coincides with the found system resonance (27Hz). Compared with its maximum values (see figure 5), the input and output signals at the resonance have about four times less power, translating in about 16 times less signal amplitude. So another possibility is that the relative contribution from the true system dynamics and noise are difficult to resolve by the linear system identification, due to the small signal to noise ratios at this frequency range. To obtain more accurate parameters we will have to resort no non-linear system identification techniques.

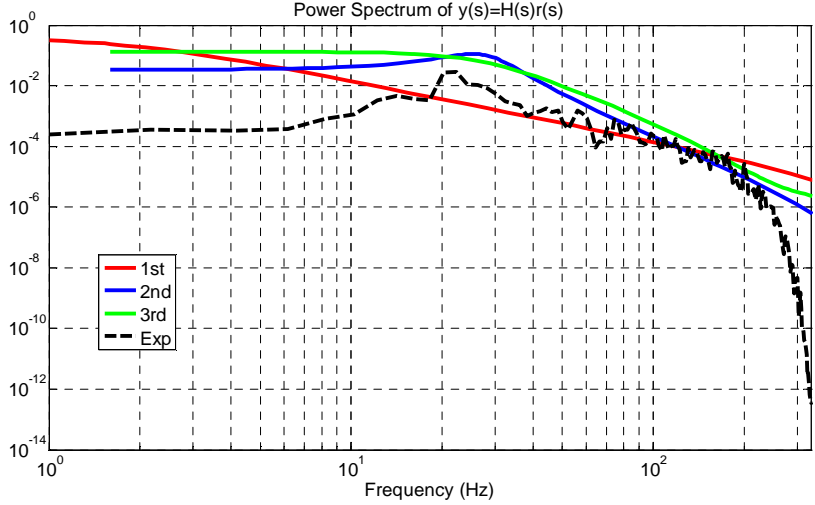

Fig. 12. Noise power spectrum for experimental data (in black) and linear models up to third order (in colour).

## VIII Non linear system identification

In this section we are going to exploit the full non-linear model to identify the parameters of the equilibrium voltage model more accurately. We will build a cost function based on the triplet $L_i, I, V_C$ and exploit the constraints imposed by the model's non-linear dependencies.

We start by using separate inputs $V_B, V_R$ and restrict the state space model to a second order structure without steady state Kalman gain. To be able to determine the most appropriate set of initial conditions, the second order approximation for equilibrium loop voltage dynamics that was found in the previous section is written in canonical observable form as

$$\begin{bmatrix} \dot{x}_1 \\ \dot{x}_2 \end{bmatrix} = \begin{bmatrix} -2\delta\omega & 1 \\ -\omega^2 & 0 \end{bmatrix} \begin{bmatrix} x_1 \\ x_2 \end{bmatrix} + \omega^2 \begin{bmatrix} kT_B & (1-k)T_R \\ k & (1-k) \end{bmatrix} \begin{bmatrix} V_B \\ V_R \end{bmatrix} \quad (58)$$

$$y = x_1$$

and $\omega, \delta$ are the natural frequency in radians per second $\delta$ damping factor. The variables have been scaled to the natural frequency so that the model parameters are all roughly of the same magnitude.

Written in this way, the corresponding transfer functions are

$$G_B(s) = \frac{V_C(s)}{V_B(s)} = \frac{k\omega^2(1+T_B s)}{s^2 + 2\delta\omega s + \omega^2} \quad (59)$$

$$G_R(s) = \frac{V_C(s)}{V_R(s)} = \frac{(1-k)\omega^2(1+T_R s)}{s^2 + 2\delta\omega s + \omega^2} \quad (60)$$

The difference with the model found in the previous section is that now we will allow separate inputs for boundary and resistive voltages. Choosing a relative weight $k \leq 1$ for $V_B$, and a relative weight $(1-k)$ for the $V_R$ input, the steady state solution of (58) is automatically guaranteed to match the steady state solution (28).

The weight $k$ regulates the relative contributions from boundary and resistive voltages to the equilibrium voltage. For instance, if $k=0.5$ both inputs account for an equal 50% contribution. This weight should be approximately in inverse relationship with the relative proximity of the equilibrium and boundary flux surfaces, according to (41).

The natural frequency $\omega$ and damping factor $\delta$ are a generalization of a "two time scale" model, which allows for resonant behavior when the damping factor is less than one. This allows us to explore new physics such as the existence of a plasma capacitance that could give resonant behavior on the loop voltage at high frequencies. As it was discussed in the previous section, this type of resonant behavior should not be observed at the sub-kHz range, but the linear system identification of the previous section has given some importance to this possibility, so we should retain it. When the damping factor is greater than one, the system has two real poles, and a typical "*two time scale dynamics*" is recovered. Regardless of the damping factor value, the natural frequency should scale in inverse relationship with the plasma temperature, just like the skin time scales in direct relationship with temperature.

The factors $T_B, T_R$, account for fast events in the high frequency end of the spectrum, so the equilibrium loop voltage knows about high frequency changes at the boundary or resistive voltages at faster rates than the resistive time scale. This accounts for processes like fast magnetic reconnection, turbulence, helicity transport [58] etc

To find the model parameters, the linear diffusion model (58) is combined with the non linear equations (16), (25)and (26) to form a closed set of equations. The input $V_R$ to the linear diffusion model (58) has now become a non linear function of the plasma current, plasma resistance and current drive, according to (16). Equations (16),(25), (26), (58) are then evolved in time starting from given initial conditions. The simulation results for both experiments (39950 and 39952) are finally compared with the observed data and a cost function is built as a weighted sum of the least squares prediction errors. A gradient descent algorithm is used to find the model parameters for which the cost function has a minimum. Since the system has outputs measured in many different units at different magnitudes, volt, μH and MA, the weighting function is chosen as the inverse of the data sample variance (50) to regularize the units and avoid excessive weighting on some of the outputs. As with any nonlinear optimization algorithm, there is a chance that the model might find a local minimum that is not accurate. To help the non linear optimization process, an initial guess for the model parameters obtained from the linear system identification analysis presented earlier is used.

The results from this non-linear optimization are shown in figure 13 (red traces), and the corresponding parameters are summarised in table 2.

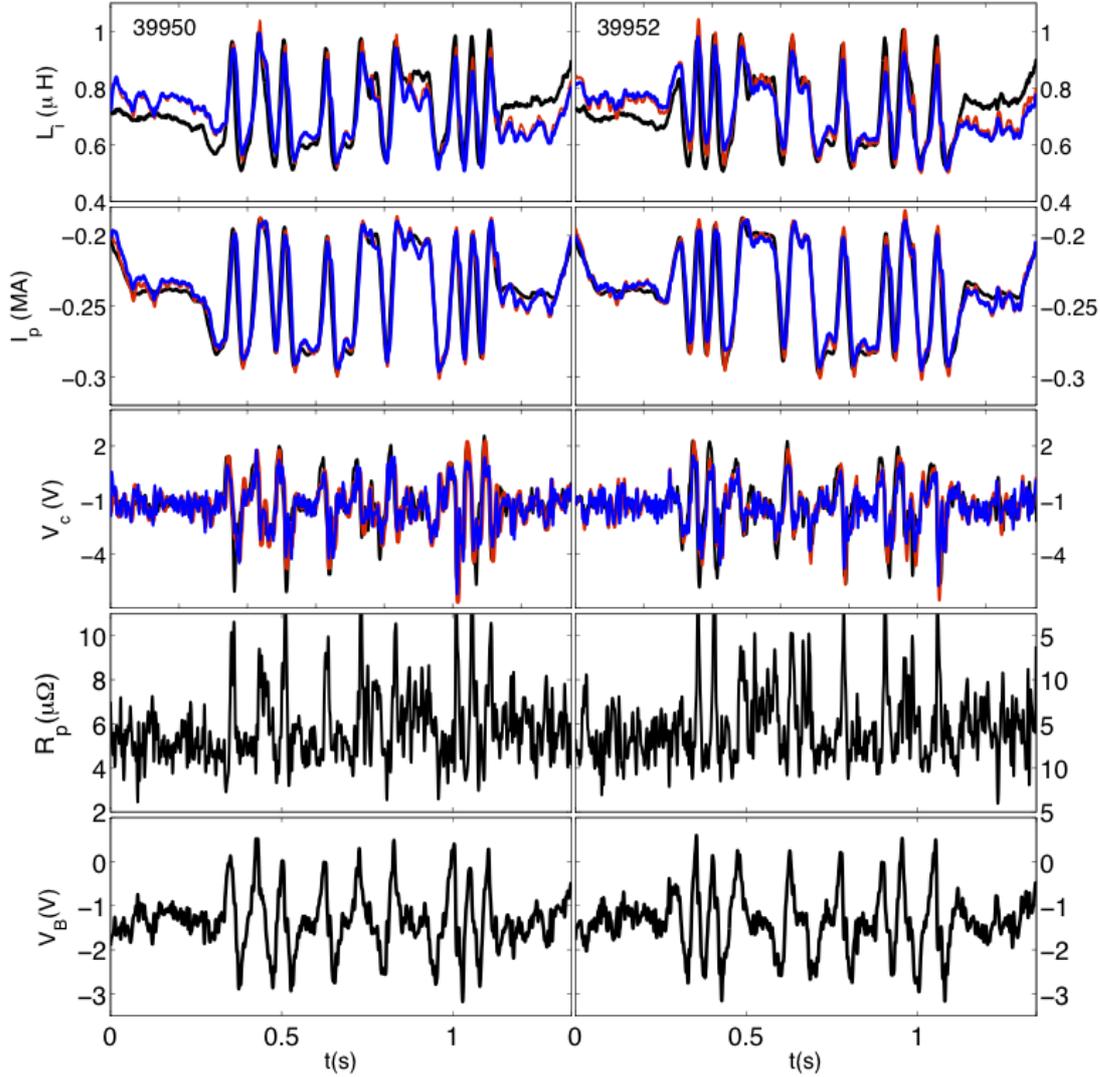

Fig. 13. Comparison between second order model simulations with an under-damped model (red) and damped model (blue). Both models produce similar results with different sets of parameters. This is an indication of a flat cost function vs. the model parameters, with broad minima.

| Shots | ω (σ) | δ (σ) | $T_B$ (σ) | $T_R$ (σ) | k (σ) |
|---|---|---|---|---|---|
| 39950-52 | 256 (1) | 0.38 (0.01) | 0.021 (0.001) | 0 (1e-5) | 0.67 (0.01) |

Table 2 . Non linear system identification using linear ($V_C$) and non linear (I, $L_i$) outputs in the objective cost function. The parameters are found for a generic second order model. The set of parameters is valid for the two shots shown.

The parameters found using the non-linear optimization correspond again with an under-damped model. We would like to exclude this result because, as mentioned earlier, resonant behaviour cannot physically exist in the sub-kHz range. So we perform a final check restricting the damping factor to be $\delta \geq 1$ in the optimization process. Both shots are identified separately both to increase the agreement with experimental data and to check parameter value dispersion. The simulation results with the optimized parameters (blue traces) are shown in figure 13 The parameters found for the two shots along with their standard deviation are given in table 3.

| Shot | ω (σ) | δ (σ) | $T_B$ (σ) | $T_R$ (σ) | k (σ) |
|---|---|---|---|---|---|
| 39950 | 439 (3) | 1 (0.04) | 0.027 (0.001) | 0 (1e-5) | 0.6 (0.01) |
| 39952 | 440 (4) | 1 (0.05) | 0.028 (0.001) | 0 (1e-5) | 0.5 (0.01) |

Table 3 Non linear system identification using linear ($V_C$) and non linear (I, $L_i$) outputs in the objective cost function. The parameters are found for a damped second order model.

Looking at the results of figure 13, it seems the damped models of Table 3 (blue traces) are also compatible with the experimental results (black traces), with small differences with respect the under damped model of Table 2 (red traces). The overall fit parameters for plasma inductance (38%), current (70%) and equilibrium voltage (30%) are almost identical for both the damped and the under damped models. This evidences a flat cost function vs. the model parameters, with broad minima.

We may inquire whether the low fit parameters obtained, particularly in the inductance and equilibrium voltage, are the result of small scale noise or represent a more fundamental failure to describe the dynamics. To this end, we will perform a correlation analysis between the model residuals and inputs. If the unexplained residuals correlate with the inputs, there are features in the outputs that can still be related to the inputs, and therefore these are unexplained features not reproduced by the model. If the correlation levels are below the confidence intervals, the discrepancy can be safely attributed to noise, and this would mean that we have reached the best possible model that can be extracted from the given data. The results can be found in figure 14. The cross correlation between the inputs and residuals for equilibrium voltage lay outside the 99% confidence intervals (dotted lines). There are features in the equilibrium voltage that correlate with inputs, and therefore are not explained by the equilibrium voltage model.

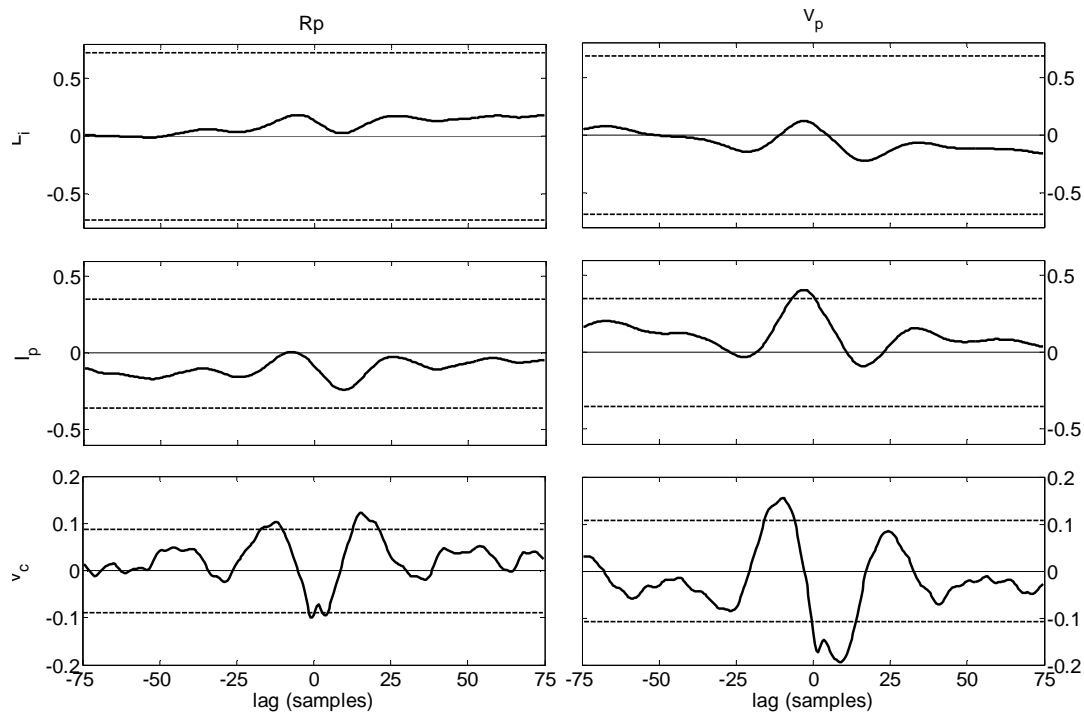

Fig. 14. Correlation function table. The left-hand column shows the correlation functions between the plasma resistance input and the residuals from plasma inductance, current and equilibrium voltage. The right-hand column shows the correlation functions between boundary loop voltage and the residuals from plasma

inductance, current and equilibrium voltage. The analysis is performed up to ±75 lags (±112ms). The discontinuous lines bound the range that has a 99% probability of being statistically insignificant.

We have no uncertainty information in the equilibrium reconstructed data, so we can not be sure about the accuracy of the second order model parameters found. Quite likely different equilibrium reconstruction methods could give different order and model parameters. Nevertheless, the lumped parameter model presented explains the most salient features of the plasma current transients, such as the inverse correlation between plasma current ramp rates and internal inductance changes, without explicit or detailed knowledge about current distribution, resistivity or loop voltage profiles. This proves that explicit or detailed knowledge of plasma profiles is not necessarily required to predict the time evolution of bulk plasma properties such as plasma inductance or current with reasonable accuracy; at least in Ohmic conditions without external heating and current drive sources. The reason for this is the smoothness of the loop voltage profile under these conditions. The skin effect transformer model provides a good balance between model accuracy and simplicity in this case.

## IX Reduced model in control affine form

We consider the case of interest where plasma resistance and non inductive current are model parameters or perturbations, and the only actuators available are the currents in the OH coil and PF systems. We define the system input, output and state vectors as

$$x = \begin{pmatrix} L_i & I & \psi_R - \psi_C & V_C \end{pmatrix}^T \tag{61}$$

$$y = \begin{pmatrix} L_i & I \end{pmatrix} \tag{62}$$

$$u = \sum_j M_j \dot{I}_j \tag{63}$$

Any combination of states can be made available in the output equation (62). In this example plasma current and internal inductance are chosen.

When the currents in the PF coil system are devoted to regulation of plasma shape and position, the plasma external and mutual inductances are stationary in time. The dynamics internal inductance (32), plasma current (33), flux and equilibrium loop voltage (58) can then be written in state space form as

$$\begin{aligned} \dot{x} &= f(x) + g(x)u \\ y &= h(x) \end{aligned} \tag{64}$$

$$f(x) = \begin{bmatrix} \dfrac{2\left(R\left(x_2 - \hat{I}\right) - x_4\right)}{x_2} \\ \dfrac{\left(x_4 - 2R\left(x_2 - \hat{I}\right)\right)}{L_e + x_1} \\ -\left(R\left(x_2 - \hat{I}\right) - x_4\right) \\ \omega^2\left((k-1)x_3 + kx_1x_2\right) - \omega^2 R\left(x_2 - \hat{I}\right)\alpha(x_1) - x_4\beta(x_1) \end{bmatrix} \tag{65}$$

$$\alpha(x_1) = \left( T_R(k-1) - \frac{2L_e T_B k}{L_e + x_1} \right)$$

$$\beta(x_1) = \left( 2\delta\omega + \frac{2L_e T_B k \omega^2}{L_e + x_1} \right) \tag{66}$$

$$g(x) = \begin{pmatrix} 0 & \dfrac{-1}{L_e + x_1} & 0 & \dfrac{-k\omega^2 T_B x_1}{L_e + x_1} \end{pmatrix}^T \tag{67}$$

$$h(x) = \begin{pmatrix} x_1 & x_2 \end{pmatrix} \tag{68}$$

The system can be trivially augmented with additional state equations to obtain additional variables or interest. For instance, writing

$$\dot{x}_5 = -x_4 \tag{69}$$

$$y = \begin{pmatrix} x_5 & x_5 + x_3 & x_5 - x_1 x_2 \end{pmatrix} \tag{70}$$

the equilibrium, resistive and boundary fluxes are available in the output vector.
Likewise, the dimensionless flux shape factor [14]

$$\chi = \frac{\psi_C - \psi_B}{\psi_R - \psi_B} \tag{71}$$

can be obtained as a combination of the system states as

$$y = \left( \frac{x_1 x_2}{x_1 x_2 + x_3} \right) \tag{72}$$

Written in the control affine form, the model is ready to be used for the design of control systems for plasma current, inductance or any other variable of interest that can be obtained as function of the state vector (61).

## X Conclusions

A lumped parameter, state space model for the tokamak transformer including the slow flux penetration in the plasma (skin effect transformer model) has been developed and validated with TCV tokamak discharges.

The model does not require detailed or explicit information about plasma profiles or geometry. Instead, this information is lumped in system variables, parameters and inputs. Exact expressions for these and fundamental relationships among them have been derived from basic electromagnetic theory.

The model has an exact mathematical structure built from energy and flux conservation theorems, predicting the evolution and non linear interaction of the plasma current and internal inductance as functions of the primary coil currents, plasma resistance, non-inductive current drive and the loop voltage at a specific location inside the plasma termed the equilibrium loop voltage.

Loop voltage profile in the plasma is substituted by a three-point discretization, and ordinary differential equations are used to predict the equilibrium loop voltage as function of the boundary and resistive loop voltages. This provides a simplified model for equilibrium loop voltage evolution, which is reminiscent of the skin effect. A systematic procedure has been applied to find the order this differential equation and its parameters in terms of objective cost functions, quantitative information criteria, frequency response, and residuals whiteness/correlation tests.

Fast plasma current modulation experiments with Random Binary Signals (RBS) have been conducted in the TCV tokamak to generate the required data for the analysis. Plasma current was modulated in Ohmic conditions between 200kA and 300kA with 30ms rise time, several times faster than its time constant $L_p/R \approx 200$ms.

A second order differential equation has been found to be the best approximation for equilibrium loop voltage model. When the equilibrium voltage model is combined with the energy and flux preserving equations, the resulting non linear model is capable to describe the plasma current and internal inductance time evolution with 70% and 38% accuracy under these modulation conditions. The skin effect transformer model explains the most salient features of the plasma current transients, such as the inverse correlation between plasma current ramp rates and internal inductance changes. This proves that explicit or detailed knowledge of plasma profiles is not necessarily required to predict the time evolution of bulk plasma properties such as plasma inductance or current with reasonable accuracy; at least in Ohmic conditions without external heating and current drive sources. The reason for this is the smoothness of the loop voltage profile under these conditions. The skin effect transformer model provides a good balance between model accuracy and simplicity in this case.

**Acknowledgements**

This work has been supported by the Euratom mobility programme, the EU FP7 EFDA under the task WP09-DIA-02-01 WP III-2-c, the Swiss National Science Foundation, the University of the Basque Country (UPV/EHU) through Research Project GIU11/02, and the Spanish Ministry of Science and Innovation (MICINN) through Research Project ENE2010-18345.
The authors acknowledge interesting discussions with J. Alonso (Ciemat).